\documentclass[twocolumn,aps,showpacs]{revtex4}
\usepackage{graphics}
\begin{document}
\title{Second generation wave-function thermostat for ab-initio
molecular dynamics}
\author{Peter E. Bl\"ochl}
\affiliation{Clausthal University of Technology, Institute for Theoretical
Physics, Leibnizstr.10, D-38678 Clausthal-Zellerfeld, Germany}
\begin{abstract}
  A rigorous two-thermostat formulation for ab-initio molecular
  dynamics using the fictitious Lagrangian approach is presented.  It
  integrates the concepts of mass renormalization and temperature
  control for the wave functions. The new thermostat adapts to the
  instantaneous kinetic energy of the nuclei and thus minimizes its
  influence on the dynamics. Deviations from the canonical ensemble,
  which are possible in the previous two-thermostat formulation, are
  avoided. The method uses a model for the effective mass of the wave
  functions, which is open to systematic improvement.
\end{abstract}
\date{\today} 
\maketitle
%
\section{Introduction}\label{sec0}
%
Ab-initio molecular dynamics\cite{Car85}, also called the
Car-Parrinello method, allows to study the dynamical and finite
temperature behavior of molecules, surfaces and solids with forces
from highly accurate first principles density functional
simulations\cite{Kohn,KohnSham}. This approach has revolutionized the
way electronic structure calculations are done to date. Two main
variants of this approach have established themselves. The original
approach of Car and Parrinello\cite{Car85} uses a fictitious
Lagrangian to deduce a dynamical equation of motion for the wave
functions, while the so-called ``exact'' Born Oppenheimer
dynamics\cite{BOdyn} performs self-consistency loops for each set of
atomic positions.

The underlying idea of the fictitious Lagrangian approach, which
treats the electronic wave functions as dynamical fields that obey a
Newton-type equation of motion. If the temperature attributed to the
motion of the wave functions is sufficiently low, the wave functions
propagate close to the Born-Oppenheimer surface (i.e. the
instantaneous electronic ground state). Thus the electronic wave
functions can be propagated with relatively minor computational effort
while the nuclei are moving.  Self-consistency loops for each step of
the trajectory are avoided.

A difficulty arises from the requirement that the wave functions
remain sufficiently close to the Born-Oppenheimer surface so that
physically correct forces on the nuclei are produced. This implies
that the wave functions must remain ``cold'', while at the same time
the nuclei are at the, relatively high, physical temperature.  Hence
ab-initio molecular dynamics simulations are in principle
non-equilibrium simulations. 

In most cases, the heat transfer between the electronic and the atomic
subsystems is sufficiently slow as a result of the separation between
electronic and nuclear frequency spectra\cite{Pastore91}.  For long
simulations, however, one has to bear in mind that the nuclear and
electronic variables will eventually, even though very slowly,
approach thermal equilibrium, where the wave functions deviate from
the Born-Oppenheimer surface and the forces acting on the atoms are
unphysical. In order to rigorously maintain a stable simulation, two
thermostats\cite{Nose84,Hoover85} are introduced\cite{Bloechl92a}. One
thermostat keeps the nuclei at their physical temperature and the
other one adsorbs the additional heat transfer to the wave
functions. The optimum temperature for the wave-function dynamics,
which is required so that the wave functions can follow the nuclei
adiabatically, has been discussed earlier.\cite{Bloechl92a}

However, the previous two-thermostat formulation has three
deficiencies:
\begin{itemize}
\item For systems consisting of parts that are only in weak thermal
contact, the thermostats do not guarantee that the canonical ensemble
is correctly sampled.  The heat transfer from the nuclei to the wave
functions is larger for some atoms than for others, which may result
in different effective temperatures for different parts of the system.
\item The target kinetic energy for the wave function dynamics is
constant over time and is determined according to the target
temperature of the atoms. For small systems the fluctuations in the
atomic kinetic energy can be sizeable. In that case the wave function
thermostat will heat the wave functions when the atomic kinetic energy
is low and cool more than necessary if the atoms have a high kinetic
energy. This has an adverse effect on the nuclear trajectories.
\item The goal of the wave function thermostat is to keep the wave
functions cold. The thermostat however induces fluctuations in the
wave function kinetic energy and thus introduces undesired heating
sequences.
\end{itemize}

In this paper, I analyze the trajectories of the nuclei and propose an
new formulation of the two-thermostat method, which overcomes these
deficiencies. The new method links the target temperature for the
electronic wave functions directly to the instantaneous motion of the
atoms.  Thus the wave function thermostat adapts to the temperature
fluctuations of the nuclei, and minimizes its influence on the
atomic trajectories.  Secondly, the indirect influence of the
wave-function thermostat on the atomic trajectories is compensated
using an additional, opposing friction term in the equations of motion
for the atoms.

A detailed numerical analysis of the errors in the forces resulting
from the deviations from the Born-Oppenheimer surface and due to the
previous two-thermostat formulation been performed\cite{Tangney}.

In Sec.~\ref{s_oldthermostat} the previous two-thermostat formulation
is restated, which allows me to introduce my notation.  In
Sec.~\ref{s_downfolding} effective equations of motion for the nuclei
under the influence of the wave function dynamics are derived.
Sec.~\ref{s_instability} discusses a potential thermodynamical
instability in the previous two thermostat formulation of
Car-Parrinello dynamics. Sec.~\ref{s_newthermostat} describes the new
two thermostat formulation. An approximate scheme for deriving the
parameters of the new theory and technicalities of the present
implementation are given in Sec.~\ref{s_implementation}. Test
calculations are presented in Sec.~\ref{s_tests}.
%
\section{Equations of motion with two Thermostats}
\label{s_oldthermostat}
%
Let me first describe my notation.  The electrons are described by
one-particle wave functions $|\Psi_n\rangle$. The wave functions
$|\Psi_n\rangle$ are related by a linear transformation
$|\Psi_n\rangle=T|\tilde\Psi_n\rangle$ to the variational parameters
$|\tilde\Psi_n\rangle$. Often the variational parameters are vectors
and the transformation is defined by a basis set $|\chi_k\rangle$ in
the form $|\Psi_n\rangle=\sum_k |\chi_k\rangle \tilde{\Psi}_{k,n}$.
Here, I have in mind the Projector-Augmented Wave (PAW)
method\cite{PAW94}, where the variational parameters are themselves
fields, namely the so-called pseudo wave functions
$|\tilde{\Psi}\rangle$. The pseudo wave functions are expanded in a
basis set, a detail that does not concern here. The pseudo
wave functions of the PAW method are conceptionally identical to the
wave functions of the pseudopotential formalism.
 
In the ab-initio molecular dynamics method with the previous
two-thermostat formulation, the following coupled system of equations
of motion is solved.
\begin{eqnarray}
  m_\Psi\vert\ddot{\tilde\Psi}_n\rangle 
   &=& - \tilde{H} \vert\tilde\Psi_n\rangle
   + \sum_{m} \tilde{O}|\tilde\Psi_m\rangle \Lambda_{m,n}
   - m_\Psi\vert\dot{\tilde\Psi}_n   \rangle \dot x_\Psi 
\nonumber\\
   M_i \ddot R_i &=& F_i - M_i \dot R_i  \dot x_R \, .
\nonumber\\
  Q_\Psi \ddot x_\Psi
  &=& 2 ( \sum_{n} \langle\dot{\tilde\Psi}_n|m_\Psi|\dot{\tilde\Psi}_n\rangle
 - E_{kin,0} )
\nonumber\\
   Q_R \ddot x_R
   &=& 2 ( \sum_{i} {1 \over 2} M_i \dot R_i ^2 - {1 \over 2} g k_B T )
   \, .
\label{eq_oldthermostat}
\end{eqnarray}
If $E[R,|\Psi\rangle]$ is the Kohn-Sham total energy functional, the
pseudo Hamiltonian $\tilde{H}$ is defined such that
$\frac{dE}{d\langle\tilde{\Psi}_n|}= \tilde{H}|\tilde{\Psi}_n\rangle$
and the forces $F_i$ are the partial derivatives of the Kohn-Sham
total energy functional with respect to the atomic positions. The
pseudo overlap operator $\tilde{O}=T^\dagger T$ is obtained from the
transformation between the true wave functions $|\Psi\rangle$ and the
pseudo wave functions $|\tilde\Psi\rangle$.  The overlap between two
wave functions is $\langle\Psi_n|\Psi_m\rangle
=\langle\tilde\Psi_n|\tilde{O}|\tilde\Psi_m\rangle$.  The Lagrange
parameters introduced to keep the wave functions orthogonal are
denoted as $\Lambda_{m,n}$.  The mass of the wave functions is
$m_\Psi$. In practice an operator diagonal in a plane wave
representation is used. The nuclear masses are denoted as $M_i$, where
i is the index of the corresponding atom.  $Q_\Psi$ and $Q_E$ are the
``masses'' of the thermostat variables $x_\Psi$ and $x_R$ for wave
functions and nuclei respectively. These masses determine the reponse
time and the dominant frequency of the thermostats. $g$ is the number
of nuclear degrees of freedom and $k_B$ is the Boltzmann
constant. $E_{kin,0}$ is a parameter that determines the target
kinetic energy of the wave functions in the simulations. Its value is
chosen according to an estimate of the Born-Oppenheimer kinetic energy
of the wave functions. In the entire paper Hartree atomic units
($\hbar=e=m_e=4\pi\epsilon_0=1$) are used. 
The present formulation of the thermostat differs from
Hoover's: the thermostat variable used by Hoover corresponds to the
{\em time derivatives} of the thermostat variables $x_R$ and $x_\Psi$
used here. This choice has the advantage that the thermostats obey
second order differential equations just as the nuclear position and
the wave functions.

Even though the equations of motion given above cannot be derived
from a Lagrangian formalism, they have a conserved energy
\begin{eqnarray}
E_{c}&=&\sum_n\langle\dot{\tilde\Psi}_n|m_\Psi|\dot{\tilde\Psi}_n\rangle
+\sum_i \frac{1}{2}M_i\dot{R}_i^2
+E[\vert\Psi_n\rangle,R_i]
\nonumber\\
&+&\frac{1}{2}Q_\Psi \dot{x}_\Psi^2+2E_{kin,0}x_\Psi
+\frac{1}{2}Q_R \dot{x}_R^2+gk_BT x_R \ .
\end{eqnarray}
The first two terms are the kinetic energies of wave functions and
nuclei, the third is the potential energy, i.e. the density-functional
total energy. The remaining terms are kinetic and potential energies
of the two thermostats for wave functions and nuclei.

Simulations with the two thermostats reach a stationary state in which
the wave function thermostat variable $x_\Psi$ exhibits an
approximately constant drift to larger values, which freezes out the
deviations from the Born-Oppenheimer approximation. As a consequence
of energy conservation\cite{footnote5} the atom thermostat variable
decreases at an average rate
\begin{equation}
  \langle\dot{x}_R\rangle=-\frac{E_{kin_0}}{\frac{1}{2}gk_BT}
  \langle\dot{x}_\Psi\rangle
\label{linkthermostats}\ , 
\end{equation}
which restores the energy absorbed by the wave-function thermostat.

\section{Effective equations of motion for the nuclei}
\label{s_downfolding}

In order to understand the motion of the nuclei with the wave
functions and thermostats tied to them, let me derive effective
equations of motions for the nuclei. It is useful to consider the
atoms as quasiparticles consisting each of a nucleus and the wave
function cloud, i.e. the distortion of the surrounding electron
gas. Just as an electron distorts a surrounding crystal lattice to
form a polaron, here a nucleus distorts the electron gas to form a
quasiparticle called an atom. The distortion of the electron gas,
respectively its wave functions, will be called wave function cloud.
The effective equations of motion are obtained from the equations of
motion in Eq.~\ref{eq_oldthermostat} by constraining the wave functions
to be identical to the exact Born-Oppenheimer wave function.  The
details of the derivation are given in App.~\ref{a_downfolding}. Here
only the result is shown:
\begin{eqnarray}
\sum_j[M_i\delta_{i,j}+K_{i,j}]\ddot{R}_j&=& F_i-M_i\dot{R}_i\dot{x}_R
-\sum_jK_{i,j}\dot{R}_j\dot{x}_\Psi
\nonumber\\
&-&\sum_{j,k} \Bigl(\frac{\partial K_{i,j}}{\partial R_k}-\frac{1}{2}\frac{\partial
  K_{j,k}}{\partial R_i}\Bigr)\dot{R}_j\dot{R}_k \ .
\label{eq_downfold}
\end{eqnarray}
$K_{i,j}$ is the effective mass tensor as function of the
nuclear positions
\begin{equation}
K_{i,j}=2\sum_n\langle 
\frac{\partial\tilde\Psi_n^{BO}}{\partial{R}_i}|m_\Psi|
\frac{\partial\tilde\Psi_n^{BO}}{\partial{R}_j}\rangle \ ,
\label{effmass}
\end{equation}
where $|\tilde\Psi_n^{BO}\rangle$ are the Born-Oppenheimer wave
functions.

Let me discuss this equation: 
\begin{itemize}
\item The nuclei obtain an effective mass
$M_i\delta_{i,j}+K_{i,j}$: The atoms appear to be heavier than the
nuclei alone, because also the wave functions need to be accelerated,
whenever the nuclei accelerate. This effect has been realized
before\cite{Ppc,PAW94} and corrections are in common
use today.  
\item The third term on the right-hand side of Eq.~\ref{eq_downfold}
is related to the friction imposed by the wave function thermostat on
the wave functions. As the thermostat removes kinetic energy from the
wave functions, also the atoms are cooled down.  In particular for two
loosely coupled subsystems, where one subsystem has a different
average effective wave function mass than the other, a drift of the
thermostat variables cools one subsystem at the cost of the other.
This may cause deviations from the canonical ensemble as discussed
in more detail in the following section.
\item The last term in Eq.~\ref{eq_downfold} describes the effect of
the changes of the effective mass as the atoms are moving around.
This term will not be discussed in more detail, because in this paper
an approximation with the effective masses are independent of the
atomic positions will be employed, where this term vanishes.
\end{itemize}
%
\section{Deviations from the canonical Ensemble}
\label{s_instability}
Here, the deviations from the canonical ensemble mentioned above are
investigated. The rate of energy change of a subsystem $A$ imposed by
the thermostats is according to Eq.~\ref{eq_downfold}
\begin{equation}
\dot{E}_A=-\sum_{i\in A} M_i
\dot{R}_i^2\dot{x}_R-\sum_{i,j\in A }K_{i,j}\dot{R_i}\dot{R}_j\dot{x}_\Psi
\label{edrift1}
\end{equation}
Assuming that subsystem $A$ is in thermal equilibrium at a temperature
$T_A$, the thermal average of $\dot{R}_i\dot{R_j}$ is
$\langle\dot{R}_i\dot{R_j}\rangle_T=\delta_{i,j}k_BT_A/M_i$.
Furthermore, the drifts of the two thermostat variables are related by
Eq.~\ref{linkthermostats}.\cite{footnote1} Thus Eq.~\ref{edrift1} can
be transformed to
\begin{eqnarray}
\langle\dot{E}_A\rangle_T
&=&g_Ak_BT_A\langle\dot{x}_\Psi\rangle 
[\frac{E_{kin,0}}{gk_BT/2}-\frac{1}{g_A}\sum_{i\in A }\frac{K_{i,i}}{M_i}]
\ ,
\end{eqnarray}
where $g$ and $g_A$ are the number of degrees of freedom of the total
system and subsystem $A$. 

If the thermal average of the Born-Oppenheimer wave function kinetic
energy,
\begin{eqnarray}
\sum_{i,j}\frac{1}{2}\langle \dot{R}_iK_{i,j}\dot{R}_j\rangle=
\frac{1}{2}k_BT\Bigl[\sum_{i\in A}\frac{K_{i,i}}{M_i}
+\sum_{i\in B}\frac{K_{i,i}}{M_i}\Bigr]\ ,
\end{eqnarray}
is chosen for $E_{kin,0}$, one obtains
\begin{eqnarray}
\langle\dot{E}_A\rangle_T
&=&\frac{g_Ag_B}{g_A+g_B}k_BT_A\dot{x}_\Psi
\nonumber\\
&\times&[\frac{1}{g_B}\sum_{i\in B }\frac{K_{i,i}}{M_i}
-\frac{1}{g_A}\sum_{i\in A }\frac{K_{i,i}}{M_i}]
\end{eqnarray}
where $g_B=g-g_A$ is the number of degrees of freedom in the subsystem
$B$ which together with subsystem $A$ makes up the complete system.

Hence, the effect of the thermostats can be described as a heat flow
from one system to the other, with a heat transport coefficient
\begin{eqnarray}
\lambda_x&=&-\frac{\langle\dot{E}_A\rangle_T
-\langle\dot{E}_B\rangle_T}{T_A-T_B}
\nonumber\\ &=&-\frac{g_Ag_B}{g_A+g_B}k_B\dot{x}_\Psi\Bigl[
\frac{1}{g_B}\sum_{i\in B }\frac{K_{i,i}}{M_i}
-\frac{1}{g_A}\sum_{i\in A }\frac{K_{i,i}}{M_i}\Bigr]
\end{eqnarray}
Note that $\lambda_x$ is negative. Thermal equilibrium is only reached
when total heat transport coefficient among subsystems is positive.
This clearly shows that there is a thermodynamical instability 
when the physical thermal coupling is smaller (in absolute values) than
the implicit coupling via the thermostats.

The conditions for this thermodynamic instability are that
\begin{itemize}
\item the drift in the wave function thermostat is sufficiently
  rapid,
\item $K_{i,i}/M_i$ is very different for the atoms of one
  subsystem as compared to the other, and
\item the thermal coupling between the systems is sufficiently small,
so that the effect of the thermostats is appreciable compared to the
physical thermal coupling.  
\end{itemize}
While these conditions rarely corroborate, already the fluctuations,
which result from a nearby instability, can render the results of a
simulations inaccurate.

\section{Equations of motion with controlled heat-back-feeding}
\label{s_newthermostat}
To cure the problems of the two-thermostat formulation, terms are
added to the equations of motion that compensate the effect of the
finite effective mass of the wave functions in the down-folded
equations of motion for the atoms. I proceed with the assumption that
the effective mass tensor of the wave functions is known. Later, I
will discuss an approximate expression to be used in practice.

I propose a new set of equations of motion, which is the main result
of this paper:
\begin{eqnarray}
  m_\Psi\vert\ddot{\tilde\Psi}_n\rangle 
   &=& - \tilde{H} \vert\tilde\Psi_n\rangle
   + \sum_{m} \tilde{O}|\tilde\Psi_m\rangle \Lambda_{m,n}
- m_\Psi\vert\dot{\tilde\Psi}_n
   \rangle \dot x_\Psi 
\nonumber\\
 \sum_j  (M_i\delta_{i,j}&-&K_{i,j}) \ddot R_j = F_i - M_i\dot{R}_i\dot{x}_R 
  +\sum_jK_{i,j}\dot{R}_j\dot{x}_\Psi
\nonumber\\
 &+&\sum_{i,j}(\frac{\partial K_{i,j}}{\partial
   R_k}-\frac{1}{2}\frac{\partial K_{j,k}}{\partial R_i})\dot{R_j}\dot{R_k}
\, .
\nonumber\\
  Q_\Psi \ddot x_\Psi
  &=& 2 \theta(\dot{x}_\Psi)
\Bigl[ \sum_{n} \langle\dot{\tilde\Psi}_n |m_\Psi| \dot{\tilde\Psi}_n \rangle 
  - \sum_{i,j}\frac{1}{2}K_{i,j}\dot{R}_i\dot{R}_j \Bigr]
\nonumber\\
   Q_R\ddot x_R
   &=& 2 (\sum_{i}\frac{1}{2}M_i\dot R_i^2-\frac{1}{2}gk_BT)
   \, .
\label{eq:newthermostat}
\end{eqnarray}
The step function $\theta(\dot{x})$ in the equation of motion for the
wave function thermostat will be discussed in detail below. Even if
the step function is removed, these equations create a stable and
energy conserving dynamics. Its role is to shut the thermostat down
for those periods of time, where the thermostat would otherwise heat
the wave functions.

The system of equations Eq.~\ref{eq:newthermostat} has a conserved
energy
\begin{eqnarray}
E_{c}&=&+\sum_i \frac{1}{2}M_i \dot{R}_i^2+E[\vert\Psi_n\rangle,R_i]
\nonumber\\
&+&\sum_n\langle\dot{\tilde\Psi}_n|m_\Psi|\dot{\tilde\Psi}_n\rangle
-\sum_{i,j} \frac{1}{2}K_{i,j} \dot{R}_i\dot{R}_j
\nonumber\\
&+&\frac{1}{2}Q_\Psi \dot{x}_\Psi^2+\frac{1}{2}Q_R \dot{x}_R^2+gk_BT x_R
\label{eq_newec}
\end{eqnarray}
The second line of Eq.~\ref{eq_newec} is the kinetic energy of the
{\em free} motion of the wave functions representing deviations from
the Born-Oppenheimer surface.

The new system of equations of motion differs in four points from the
ones used previously.
\begin{itemize}
\item The nuclear mass is reduced by the effective wave function mass,
so that the effective mass of atoms, namely the sum of the reduced
mass of the nuclei and mass of the wave function cloud add up to the
true nuclear mass. This correction has been previously
suggested\cite{Ppc,PAW94} and is common practice in state-of-the-art
ab-initio molecular dynamics simulations.
\item A friction force $\sum_jK_{i,j}\dot{R}_j\dot{x}_\Psi$ acting on
the atoms has been added. This term is controlled by the wave function
thermostat and opposes the drag from the wave functions, which
themselves are slowed down by the wave function thermostat.
  
  The rate of energy added to or removed from the physical system by
  the wave functions thermostat is
  $[\sum_n\langle\dot{\tilde\Psi}_n|m_\psi|\dot{\tilde\Psi}_n\rangle
  -\sum_{i,j}K_{i,j}\dot{R}_i\dot{R}_j]\dot{x}_\Psi$. This energy
  transfer is proportional to the deviation of the wave functions from
  the Born Oppenheimer surface. Hence the thermostat acts only on the
  \textit{free} oscillations of the wave functions.
  
\item The estimate $\frac{1}{2}\sum_{i,j}K_{i,j}\dot{R}_i\dot{R}_j$
  for the instantaneous Born-Oppenheimer kinetic energy has been
  introduced into the feedback equations for the wave function
  thermostat instead of a constant target energy used previously. This
  modification was necessary in order to obtain a conserved energy.
  
  Besides restoring energy conservation, this term has another
  beneficial role, which is most important for systems with large
  kinetic energy fluctuations. The wave function thermostat adapts to
  the instantaneous kinetic energy of the nuclei and acts only on the
  deviations of the wave functions from the Born-Oppenheimer surface:
  The wave function thermostat would remain inactive for
  Born-Oppenheimer wave functions.
\item The wave function thermostat shuts down in an energy conserving
fashion when it would otherwise heat the wave functions. This is
discussed in the following.
\end{itemize}

Despite the more complex equations of motion the conserved
energy has one term less. There is no potential energy term related to
the wave functions, which has important implications. The thermostat
variable oscillates repeatedly between cooling and heating the wave
functions. Heating the wave functions is not desirable, as one would
like to keep them at their lowest temperature compatible with the
dynamics.

A better strategy is to couple the wave function thermostat to the
system only when cooling is required. This can be done in an energy
conserving way, since the contribution of the wave function thermostat
to the conserved energy vanishes whenever $\dot{x}_\Psi$ vanishes.
Thus the equation of motion for the wave function thermostat in
Eq.~\ref{eq:newthermostat} has been modified by introducing a step
function $\theta(\dot{x}_\Psi)$. $\theta(x)$ is the Heaviside step
function defined as $\theta(x\ge0)=1$, $\theta(0)=\frac{1}{2}$ and
$\theta(x<0)=0$. The discretization of the equation of motion for the
wave function thermostat including the step function is not
straightforward.  This equation means the following: The thermostat
dynamics is switched on with zero velocity, when the wave function
kinetic energy grows larger than the target. $\dot{x}_\Psi$ grows and
transfers energy from the wave function dynamics into the nuclear
subsystem until the velocity of the thermostat variable vanishes
again. At this point the wave function kinetic energy is below its
target. Instead of allowing the thermostat to heat the system, the
thermostat variable $x_\Psi$ is instead kept constant, implying that
the thermostat is effectively switched off.  If the equation of motion
is solved continuously, the velocity of the thermostat vanishes
exactly, when the thermostat shuts down. The velocity will remain
zero, until the wave function kinetic energy grows above its target,
and thus does not affect the dynamics for that period of time.  The
thermostat switches on again, when the kinetic energy grows above the
target.  Thus the trajectories proceed without any perturbation as long
the wave function kinetic energy remains below its target value. In a
discretized equation of motion the velocity of the thermostat must be
explicitely reset to zero, whenever it would otherwise turn negative.

To summarize, the thermostat resets the kinetic energy to a lower
value if the wave functions become too hot, while transfering the
energy into the nuclear dynamics. Note that even when the wave
function thermostat is on, the atomic trajectories are affected only
if the effective mass tensor is inaccurate. Hence the thermostat can
operate more strongly before affecting the atomic dynamics.

The thermostat variable of the atom thermostat does not experience any
longer a steady drift in the new formulation. This can be deduced from
the conserved energy expression as follows: A drift of the thermostat
variables is possible in principle, because a fixed translation of a
thermostat variable does not affect the dynamics and is therefore not
observable. The time averaged drift of $x_R$ 
vanishes, because the potential energy of the atom thermostat is
the only term in the conserved energy that depends on the thermostat
variable. All other terms in the conserved energy are observable
and therefore stationary. 

A drift of the wave-function thermostat variable does not affect the
conserved energy. Hence, a drift of $x_\Psi$ is possible.  This drift
counter-balances the heat flow from the nuclei to the wave functions
but does not add to the conserved energy.

In the remainder of this section I will discuss a possible dynamical
instability and how it can be avoided. The Born-Oppenheimer kinetic
energy enters with a negative sign, which may point to a possible
instability of the wave functions, when the bare mass-tensor with
elements $M_i\delta_{i,j}-K_{i,j}$ turns out not to be positive
definite: A particle with mass accelerates in the opposite direction
of the forces, and runs away from a minimum. The condition for the
bare mass tensor to be positive definite poses a strict upper bound on
the wave function mass $m_\Psi$.

In practice one needs to be even more restrictive: There is an
internal excitation of the quasiparticle ``atom'', where the nucleus
oscillates with high frequency about the wave function cloud. This
mode can be identified clearly when the wave functions are frozen and
only the bare nuclei are allowed to move. If the bare mass of the
nuclei is small, these oscillations may have high frequency that
require a small time step in the discretization. While this mode falls
into the class of deviations from the Born-Oppenheimer surface, the
thermostat does not cure the problem since the wave function kinetic
energy is smaller than its target
$\frac{1}{2}\sum_{i,j}K_{i,j}\dot{R}_i\dot{R}_j$. To avoid that
problem I suggest to keep the wave function mass $m_\Psi$ sufficiently
small, so that $K_{i,i}<\frac{1}{2}M_i$.

\section{Implementation}
\label{s_implementation}
\subsection{The model of an infinitely dilute gas}

The new equations of motion require an analytic expression that
approximates the effective mass tensor. As a start, I adopt the model
of an infinitely dilute gas of atoms, which has been used in
the context of the previous wave function thermostat and mass
renormalization.  The model of an infinitely dilute gas assumes that
the Born-Oppenheimer wave functions can be divided up into purely atomic
contributions, and that those are identical to the wave functions of
the corresponding isolated atoms.

The rationale for the model of an infinitely dilute gas is that the
most rapid variations of the wave function occur near the nucleus, and
are little affected by the bonding environment. Thus, already the
isolated atom will capture most of the relevant contributions.  One
can envisage better approximations for the effective mass of the wave
functions, which depend explicitely on the atomic positions, and which
may be derived from a tight-binding like description.

I insert the atomic wave function into the expression Eq.~\ref{effmass} for
the effective mass tensor of a given atom and obtain
\begin{equation}
K_{i,j}=-2\sum_n\langle\tilde\Psi_{n}|\nabla_i
m_\Psi\nabla_j|\tilde\Psi_{n}\rangle
\end{equation}

It can be shown that in this approximation $K_{i,j}$ is diagonal for
each atom with three identical matrix elements on its main diagonal:
For symmetry reasons the acceleration of an isotropic atom is parallel
to the force, that is $F=\lambda \ddot{R}$ with some parameter
$\lambda$.  Hence, the acceleration is an eigenvector of the effective
mass tensor $(M\delta_{i,j}+K_{i,j}) \ddot{R}_j=\lambda\ddot{R}_j$.
Since $\lambda$ is independent of the direction of the applied force,
$K$ has three identical eigenvalues. Therefore $K_{i,j}$ has the form
of a unity matrix times a constant and the identity
$K_{i,j}=\delta_{i,j}\frac{1}{3}{\rm Tr}K$ holds.

Thus one can obtain a more simple form of our model effective mass as
\begin{equation}
K_{i,j}=\frac{2}{3}\delta_{i,j}\sum_n\langle\tilde\Psi_{n}|-\sum_i \nabla_i
m_\Psi\nabla_i|\tilde\Psi_{n}\rangle
\end{equation}

In the special case of a G-independent mass\cite{footnote2}, the
weight of the wave function cloud is directly related to the kinetic
energy of the pseudo wave functions, as reported
previously\cite{Bloechl92a}.
\begin{equation}
K_{i,j}=\frac{4m_\Psi}{3}\delta_{i,j}\sum_n\langle\tilde\Psi_{n}|-\frac{1}{2}\nabla^2
|\tilde\Psi_{n}\rangle
\end{equation}

A G-dependent wave function mass, usually a mass-tensor diagonal in
reciprocal space with elements depending on the reciprocal space
vector of the augmented plane waves, is nowadays common practice.  It
allows to control the rapid oscillations of plane waves with large
wave-vectors\cite{Tassone94} and avoids instabilities that occur otherwise when
the basis set is increased. There are several choices for the
G-dependence of the wave function mass.  I use an expression for the
effective mass
\begin{equation}
m_\Psi(G,G')=m_\Psi^0(1+cG^2)\delta_{G,G'}
\end{equation}
In order to obtain the effective mass tensor, I determine the 
pseudo wave functions of the atom and transform them into G-space
via a Bessel transform. Then I evaluate
\begin{eqnarray}
A&=&\sum_n f_n \int dG G^2 |\tilde\Psi_n(G)|^2
\\
B&=&\sum_n f_n \int dG G^4 |\tilde\Psi_n(G)|^2
\end{eqnarray}
which are combined with the chosen parameters for the wave function
mass to the effective mass tensor of the wave functions 
\begin{equation}
K_{i,j}=\delta_{i,j}\frac{2}{3}m_\Psi^0(A_i+cB_i).
\label{eq:effmass}
\end{equation}
The variables $f_n$ are the occupation numbers of the one-particle
states.  Typical values for $A_i$ and $B_i$ are given in
Tab.~\ref{mparam}.  Note, that these values depend on the choice of
pseudo wave functions and are not transferable. They are listed here
solely to provide the reader with a feeling of the order of magnitudes
involved.
\begin{table}[htbp]
  \begin{tabular}{|l|c|c|c|c|}
\hline
Atom & $A$ & $B$ & $10^4\frac{A}{M}$ & $10^4\frac{B}{M}$ \\
\hline
H  &  0.830 &  2.830 & 4.518 & 15.403\\
He &  3.505 & 17.128 & 4.803 & 23.475\\
C  &  5.837 & 22.292 & 2.666 & 10.181\\
O  & 14.569 & 86.109 & 4.995 & 29.525\\
F  & 20.304 &143.785 & 5.863 & 41.518\\
Si &  3.327 &  6.320 & 0.650 &  1.234\\
Cl & 12.821 & 41.841 & 1.984 &  6.474\\
Fe & 10.950 & 71.366 & 1.075 &  7.010\\
Ru & 17.606 & 95.952 & 0.956 &  5.208\\
Os & 16.874 & 86.002 & 0.848 &  4.320\\
\hline
  \end{tabular}
\caption{Coefficients $A$ and $B$ for Eq.\ref{eq:effmass} for
different elements and their values relative to the nuclear masses.
These values are not transferable.}
\label{mparam}
\end{table}

The effective wave function mass must be substantially smaller than
the nuclear masses, i.e.  $\frac{2}{3}m_\Psi(A+cB)<M$, in order to obtain a
stable dynamics. If this requirement is violated the reduced mass of
the atom is negative, and atoms accelerate in the opposite
direction of the force acting on them.

\subsection{Discretized equations of motion}

All equations of motion are implemented using the Verlet algorithm. An
equation of motion for a general coordinate $x$ has the form
\begin{equation}
m\ddot{x}=F-m\dot{x}f \,
\end{equation}
where $f$ is a friction coefficient, which may be constant or imposed
by a thermostat. The thermostat $x$ creates a
canonical ensemble by tuning the friction via $f=\dot{x}$ with time.

The equations of motion discretized with a time step $\Delta$ are
obtained by replacing derivatives by the differential quotients
$\dot{x}=(x(t+\Delta)-x(t-\Delta))/(2\Delta)+O(\Delta^2)$ and
$\ddot{x}=(x(t+\Delta)-2x(t)+x(t-\Delta))/\Delta^2+O(\Delta^2)$ as
\begin{equation}
x(t+\Delta)=\frac{2}{1+a}x(t)-\frac{1-a}{1+a}x(t-\Delta)
+F(t)\frac{\Delta^2}{m}\frac{1}{1+a}
\end{equation}
where $a=f\Delta/2$.

The choice $a=0$ yields energy conserving trajectories and $a=1$
results in steepest descent dynamics. Intermediate values of $a$ are
the regime of friction dynamics. 

Because the thermostat can be propagated only with the knowledge of
the instantaneous kinetic energy, which in turn depends on the
propagated value of the thermostat, I extrapolate the thermostat
variable for $t+\Delta$ from the present and the previous two
thermostat values
\begin{eqnarray}
x(t+\Delta)&=&4x(t)-6x(t-\Delta)
\nonumber\\
&+&4x(t-2\Delta)-x(t-3\Delta)+O(\Delta^4)\ .
\end{eqnarray}
This expression introduces errors of the forces of order $\Delta^2$,
which is consistent with the overall accuracy of the Verlet algorithm.

The step function for the thermostat is implemented by resetting
$\dot{x}_\Psi$ to zero for the propagation of wave function and
nuclear positions, whenever it has a negative value otherwise, and by
setting $x_\Psi(t-n\Delta)=x(t)$, whenever the velocity
$\dot{x}_\Psi(t)$ would become negative during the propagation of the
thermostat variable.

\section{Tests}
\label{s_tests}

In order to test stability and accuracy of the method I investigated
two systems. A simulation of carbon monoxide shall illustrate how
the thermostats adapt to large fluctuations of the nuclear kinetic
energy. This test case allows also a test of the accuracy. Iron has
been used as example for a metal and shows the stability of the
trajectories against frequent band crossings.

The simulations described in the following have been performed with
the Perdew-Burke-Ernzerhof density functional\cite{PW,PBE}. The plane wave
cutoff for the wave functions has been set at 30~Ry and the one for
the density at 60~Ry. A time step of $\Delta=10$~a.u.$=0.12$~fs has
been used together with a G-independent wave function mass of
1000~m$_e$. The frequency of the thermostat for the nuclei has
been chosen to 10~THz, which is unusually large. The frequency of the
wave function thermostat has been set to 100~THz.

\subsection{Carbon Monoxide}
The unit cell size is a 13~\AA\ fcc unit cell. The electrostatic
interaction between periodic images has been subtracted.\cite{Decoupl}
A 1$\times$s+1$\times$p+1$\times$d projector set has been used, that
is one projector for every relevant set of angular momentum quantum
numbers.  The one-center density has been expanded into spherical
harmonics up to angular momentum of $\ell=2$. Parameter $A=5.87$ for
carbon and $A=15.42$ for oxygen have been used, resulting in an
effective mass of the wave function cloud amounts to 17~\% and 35~\%
of the physical nuclear mass for carbon and oxygen respectively.
Rotations and translations have been frozen out by constraints so that
a truely one-dimensional system is studied. The simulation has been
performed at 1000~K. 

The experimental value for the CO stretch vibration is
2170~cm$^{-1}$.\cite{Rao48}.The experimental bond-length is
1.1283~\AA.\cite{huber} Fixed point calculation predict a bond-length
of 1.138~\AA\ and a stretch frequency of 2125~cm$^{-1}$. These deviations
are in the range of errors expected for the density functional used.

\begin{figure}[htbp]
\resizebox{8.5cm}{!}{\includegraphics{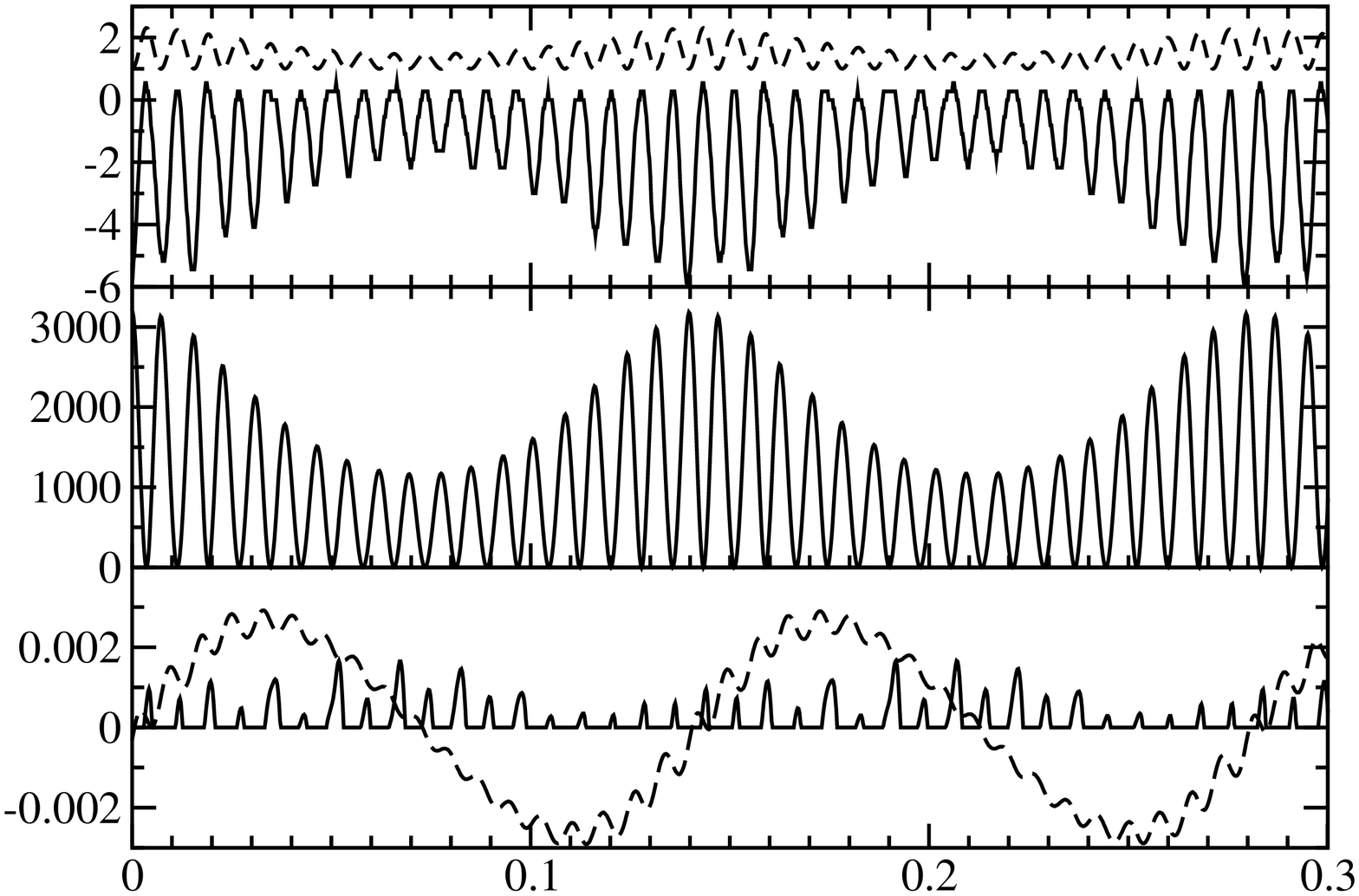}}
\caption{Energies of carbon monoxide versus time in ps.  Top:
non-Born-Oppenheimer kinetic energy of the wave functions in meV (full
line) and total energy in units of 0.1eV (dashed line) displaced
vertically. Middle: instantaneous ``temperature'' in Kelvin. Bottom:
friction imposed by the thermostats in units of $2/\Delta$ for
wave functions (full line) and nuclei (dashed line).}
\label{fig1}
\end{figure}

Simulations lasting several picoseconds have been
performed.\cite{footnote10} A sequence of 0.3~ps is shown in
Fig.~\ref{fig1}. The deviation of the wave-function kinetic energy
from the Born-Oppenheimer surface is -0.05 times the variation of the
potential energy and overall smaller than 5~meV. This indicates that
the dilute atomic gas model overestimates the effective masses by
about 10\%.  On the other hand 90~\% of the error has been removed.

The reduced mass of atoms with the wave function cloud is 25~\% of the
true reduced mass, which, without correction, would result in an
overestimate of 10\% in the frequencies. Given the error in the
effective masses one can expect frequencies that overestimate the
frequency derived from static calculation by 1~\%, which is good
agreement with the 2148~cm$^{-1}$ obtained from averaging vibrational
periods during five picoseconds.

The friction measured in units of $\frac{\Delta}{2}$ remains below
$4\times 10^{-3}\frac{2}{\Delta}$, which is very small indicating that
the heat transfer from wave functions to the nuclei is small in this
system. The total energy is conserved to within 5~meV/ps and scales
down with the size of the time step.

\subsection{Iron}
As a test of a metallic system I have chosen $\gamma$-iron.  An 8 atom
fcc unit cell has been used. Since only the $\Gamma$-point was
included in the k-point sampling, I do not expect the simulation to be
a realistic description of the material. The simulation temperature is
1185~K, at the martensitic phase transformation temperature. The
parameter $A$ was about 12~\% larger than the kinetic energy per atom,
in order to account for the promotion of s- to d-electrons as one goes
from an atom to the solid.  The effective mass of the wave function
per atom is 4.75~a.m.u (a.m.u=M($^{12}C$)/12), about 9~\% of the
nuclear mass. The atom thermostat had a period of 0.1~ps, and the wave
function thermostat had a period of 0.01~fs. The band gap due to
finite k-point sampling is typically about 0.1-0.2~eV.

\begin{figure}[htbp]
\resizebox{8.5cm}{!}{\includegraphics{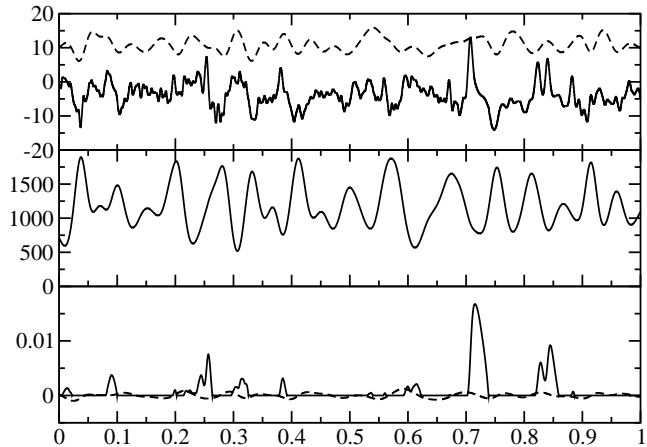}}
\caption{Energies of austenite (fcc-Iron) in an 8 atom supercell
versus time in ps.  Top: total non-Born-Oppenheimer kinetic energy of
the wave functions in meV (full line) and total energy in units of
0.1~eV (dashed). Middle: instantaneous ``temperature'' in
Kelvin. Bottom: friction imposed by the thermostats in units of
$2/\Delta$ for wave functions (full line) and nuclei (dashed line).}
\label{fig2}
\end{figure}

The results for the simulation of one picosecond is shown in
Fig.~\ref{fig2}.  The total energy variation is of order 1~eV. The
mean average kinetic energy related to the non-Born Oppenheimer motion
is 3~meV.  The total energy drifts with 0.47~meV/ps. The typical
deviation from the Born-Oppenheimer surface is 10-15~meV.
 
Most of the time the thermostat is switched off as the kinetic energy
remains below the target defined by the effective masses.  The most
pronounced quenching sequence occurs at 0.7~ps. Here the sharp
increase of the wave function kinetic energy indicates a band
crossing. A band crossing results in a randomization of the wave
functions as the occupied state changes its character into that of the
formerly unoccupied state, and would otherwise render the remainder of
the simulation useless. The wave function thermostat brings the wave
functions back to the Born Oppenheimer surface, while the perturbation
of the nuclear dynamics during this rather strong quench is minimized
by the opposing force acting on the nuclei.

\section{Conclusion}
In summary, a new formulation of the two thermostat approach for
ab-initio molecular dynamics has been presented. The approach aims at
controlling only the deviations from the Born-Oppenheimer wave
functions. The influence on the Born-Oppenheimer motion of the wave
functions and the nuclear motion is minimized by additional forces
opposing the indirect friction of the atoms. Furthermore the
thermostat is active only if the wave function kinetic energy grows
beyond its estimated Born-Oppenheimer value.  The new two-thermostat
formulation can be applied to small systems with large fluctuations of
the nuclear kinetic energy and the fictitious Born-Oppenheimer wave
function kinetic energy.

The new approach rests on an expression for the effective mass
tensor of the wave functions. A simple formula has been derived from
the previously employed model of an infinitely dilute gas. Systematic
improvements of effective mass tensor, which will improve the quality
of the simulation, can be envisaged.

\acknowledgments
Part of this work has been performed at the IBM Zurich Research
Laboratory. I wish to acknowledge the hospitality of Prof. K. Schwarz
and his group at Vienna University of Technology, where part of the
work has been performed.  This work has benefited from the
collaboration within the ESF Programme on "Electronic Structure
Calculations for Elucidating the Complex Atomistic Behavior of Solids
and Surfaces".

\appendix
\section{Downfolding the wave function dynamics}
\label{a_downfolding}
Here, the effective equations of motion Eq.~\ref{eq_downfold} for
the nuclei are derived, which include the forces of the wave
functions acting on the nuclei.

The starting point is the following set of equations
\begin{eqnarray}
m_\Psi|\ddot{\tilde\Psi}_n\rangle&=&-\tilde{H}|\tilde\Psi_n\rangle
+\tilde{O}|\tilde\Psi_m\rangle
-m_\Psi|\dot{\tilde\Psi}_n\rangle\dot{x}_\Psi
\nonumber\\
M\ddot{R}_i&=&F\ .
\end{eqnarray}

The wave functions are constrained to remain exactly on the
Born-Oppenheimer surface, 
\begin{equation}
\tilde\Psi_n(r,t)=\tilde\Psi^{BO}_n(r,R(t))\ .
\end{equation}
The Born-Oppenheimer wave functions $\tilde\Psi^{BO}_n(r,R_i)$ are the
ground state wave function for a given set of atomic positions $R_i$.

Because the Born-Oppenheimer wave functions depend on the nuclear
positions, the forces acting on the wave functions translate into
additional forces acting on the nuclei. Thus effective equations of
motion for the atoms are obtained.  The atoms are now
``quasiparticles'' consisting of nuclei and the wave function clouds
following them.

The constraints are enforced by the method of Lagrange multipliers:
The constraint forces, which describe the effect of the wave function
cloud, are the derivatives of a ``constraint energy''
\begin{equation}
E^c=\sum_n(\langle\tilde\Psi_n-\tilde\Psi^{BO}_n|\Phi_n\rangle
+\langle\Phi_n|\tilde\Psi_n-\tilde\Psi^{BO}_n\rangle)
\end{equation}
with the auxiliary fields $\Phi_n$ acting as Lagrange parameters. The
resulting constraint forces are
\begin{eqnarray}
F^c_{|\Psi_n\rangle}&=-\frac{\partial E^c}{\partial\langle\tilde\Psi_n|}
&=|\Phi_n\rangle
\\
F^c_{\langle\Psi_n|}&=-\frac{\partial E^c}{\partial|\tilde\Psi_n\rangle}
&=\langle\Phi_n|
\\
F^c_{R_i} &=-\frac{\partial E^c}{\partial R_i}
&=\sum_n(\langle\frac{\partial\tilde\Psi^{BO}_n}{\partial R_i}|\Phi_n\rangle
+\langle\Phi_n|\frac{\partial\tilde\Psi^{BO}_n}{\partial R_i}\rangle)\ .
\label{eq_fc}
\end{eqnarray}

The constraint forces are inserted in the equation of motion for
the wave functions 
\begin{equation}
m_\Psi|\ddot{\tilde\Psi}_n\rangle =
-\tilde{H}|\tilde{\Psi}_n\rangle+\sum_m\tilde{O}|\tilde\Psi_m\rangle\Lambda_{m,n}
-m_\Psi|\dot{\tilde\Psi}_n\rangle\dot{x}_\Psi -|\Phi_n\rangle\ .
\label{eq_constraintfield}
\end{equation}
With the help of the constraint condition
$|\tilde\Psi(t)\rangle=|\tilde\Psi^{BO}_n(R(t))\rangle$ and the fact
that $\tilde{H}|\tilde\Psi^{BO}_n\rangle
=\sum_m\tilde{O}|\tilde\Psi^{BO}_m\rangle\Lambda_{m,n}$ the auxiliary
fields $|\Phi_n\rangle$ are related to the Born-Oppenheimer wave
functions via Eq.~\ref{eq_constraintfield} as
\begin{eqnarray}
|\Phi_n\rangle&=&-m_\Psi|\ddot\Psi^{BO}_n\rangle
-m_\Psi|\dot\Psi^{BO}_n\rangle\dot{x}_\Psi
\nonumber\\
&=&-\sum_{i,j}m_\Psi|\frac{\partial^2\Psi^{BO}_n}{\partial R_i\partial
  R_j}\rangle\dot{R}_i\dot{R}_j
\nonumber\\
&-&\sum_i m_\Psi|\frac{\partial\Psi^{BO}_n}{\partial
R_i}\rangle\ddot{R}_i -\sum_i
m_\Psi|\frac{\partial\Psi^{BO}_n}{\partial
R_i}\rangle\dot{R}_i\dot{x}_\Psi\ .
\end{eqnarray}

The auxiliary fields are inserted into the expression for the
constraint forces acting on the atoms, Eq.~\ref{eq_fc}, and after a
few transformations the additional forces of the wave function cloud
acting on the atoms are obtained:
\begin{eqnarray}
F^c_{R_i} &=&
-\sum_n\Bigl[
\langle\frac{\partial\tilde\Psi^{BO}_n}{\partial R_i}|m_\Psi
(|\ddot{\tilde\Psi}_n^{BO}\rangle+|\dot{\tilde\Psi}_n^{BO}\rangle\dot{x}_\Psi)
\nonumber\\
&+&
(\langle\ddot{\tilde\Psi}_n^{BO}|
+\dot{x}_\Psi\langle\dot{\tilde\Psi}_n^{BO}|)m_\Psi 
|\frac{\partial\tilde\Psi^{BO}_n}{\partial R_i}\rangle)
\Bigr]
\nonumber\\
&=&-\sum_j K_{i,j}\ddot{R}_j
-\sum_{j,k} \Bigl(\frac{\partial K_{i,j}}{\partial R_k}-\frac{1}{2}\frac{\partial
  K_{j,k}}{\partial R_i}\Bigr)\dot{R}_j\dot{R}_k
\nonumber\\
&-&\sum_jK_{i,j}\dot{R}_j\dot{x}_\Psi\ .
\end{eqnarray}
The effective equation of motion given in Eq.~\ref{eq_downfold} for
the nuclei is obtained by adding the corresponding constraint forces
to the equations of motion for the nuclei.


%

\begin{thebibliography}{99}
\bibitem{footnote1} Note that this hold true only if the system
  remains in thermal equilibrium, because an ensemble
  average and a time average is mixed.
\bibitem{footnote2}A G-independent mass has the form of a constant
  times the identity operator.
\bibitem{footnote10}The simulation is not fully equilibrated, as the
thermal fluctuations are still larger than in a canonical ensemble.
This is however not relevant for the present purpose.
\end{thebibliography}
\end{document}